\documentclass[10pt,journal,compsoc]{IEEEtran}
\ifCLASSOPTIONcompsoc
  \usepackage[nocompress]{cite}
\else
  \usepackage{cite}
\fi
\usepackage{graphicx}
\usepackage{float}
\usepackage{subfigure}

\ifCLASSINFOpdf
\else
\fi
\begin{document}
\title{Gaze Gestures and Their Applications in Human-Computer Interaction with a Head-Mounted Display}

\author{W.X. Chen, 
        X.Y. Cui,~\IEEEmembership{Member,~IEEE, } 
        J. Zheng, 
        J.M. Zhang,
        S. Chen, 
        and Y.D. Yao, ~\IEEEmembership{Fellow,~IEEE}% <-this % stops a space

\IEEEcompsocitemizethanks{\IEEEcompsocthanksitem The authors are with Northeastern Univ, college of Med. \& Bio. Information Engineering, Shenyang 110004, Liaoning, Peoples R China.\protect\\
E-mail: cuixy@bmie.neu.edu.cn

% \IEEEcompsocthanksitem W.X. Chen, Y.D. Yao, J. Zhen and J.M. Zhang are with Anonymous University.
}% <-this % stops an unwanted space

% \thanks{Manuscript received April 19, 2005; revised August 26, 2015.}

}

\IEEEtitleabstractindextext{%
\begin{abstract}
A head-mounted display (HMD) is a portable and interactive display device. With the development of 5G technology, it may become a general-purpose computing platform in the future. Human-computer interaction (HCI) technology for HMDs has also been of significant interest in recent years. In addition to tracking gestures and speech, tracking human eyes as a means of interaction is highly effective. In this paper, we propose two UnityEyes-based convolutional neural network models, UEGazeNet and UEGazeNet*, which can be used for input images with low resolution and high resolution, respectively. These models can perform rapid interactions by classifying gaze trajectories (GTs), and a GTgestures dataset containing data for 10,200 "eye-painting gestures" collected from 15 individuals is established with our gaze-tracking method. We evaluated the performance both indoors and outdoors and the UEGazeNet can obtaine results 52\% and 67\% better than those of state-of-the-art networks. The generalizability of our GTgestures dataset using a variety of gaze-tracking models is evaluated, and an average recognition rate of 96.71\% is obtained by our method.
\end{abstract}

}

\maketitle

\IEEEdisplaynontitleabstractindextext

% \fi
\IEEEpeerreviewmaketitle

\IEEEraisesectionheading{
\section{Introduction}\label{sec:introduction}}

\IEEEPARstart{A} head-mounted display (HMD) is a type of computer display worn on the head or built into a helmet. Virtual reality (VR), augmented reality (AR) and mixed reality (MR) are the main applications that use HMDs. Early HMD studies were centered primarily on military applications; however, in recent years, as the cost and size of the hardware has continually decreased, this type of technology has been applied in fields such as medicine [1], education [2], industrial design [3] and entertainment [4]. An HMD is different from a traditional monitor; thus, creating appropriate forms of human-computer interaction (HCI) for HMDs is also of concern. At present, HCIs have been well established for gestures and voice input such as Microsoft’s HoloLens [5] and Magic Leap; however, these HCI methods are unsuitable when both hands are occupied or in environments in which speech is not an option. Thus, a simpler and more effective method to approach HCI with HMDs is crucial.

Eye tracking is a technique for measuring the gaze point of human eyes and their degree of movement relative to the head pose. The main task is to determine where a human is looking and for how long. The world's first noninvasive eye tracker, developed in Chicago in 1922 by Guy Thomas Buswell [6], used beams reflected from the eyes and recorded them on films to determine the gaze direction. In the 1970s, eye-tracking research advanced rapidly, especially in the field of reading research [7]. Eye tracking has been used to solve HCI problems since the 1980s [8]. 

Until now, eye-tracking applications have mainly concentrated on behavior analysis and HCI [9]. In terms of behavior analysis, by analyzing human gaze time and changes in gaze angle, we can analyze hand-eye coordination [10], students' attention in class [11], visual fatigue [12], and even emotional state [13]. In addition, eye tracking plays an auxiliary role in the diagnosis of  diseases such as autism [14], visual memory impairment [15], and mild amnestic cognitive impairment [16]. Regarding interaction, because humans can freely control their eye movements, eye-tracking technology can be used as a method for HCI. For example, gaze duration has been used to determine whether a human wants to press a button on the screen or to click a pointer on the screen via eye movements [17], [18], [19], [20], [21], [22], [23]. In recent years, HCI research based on gaze gestures has emerged. In this field, eye-tracking data are used to delineate virtual gestures that could be widely applied to HCI for games [25] and medical operations [24], [26], [27].

This paper addresses the issue of achieving HMD-based gaze interaction using an inexpensive webcam to detect and track the human gaze direction in real time at a close range and to analyze the user’s intent based on gaze trajectory data. In previous studies, the fully connected layer of the CNN network has always been used to directly detect 3D gaze coordinates [28] or two rotation vectors that represent the yaw and pitch of the gaze direction [29]. This method usually requires the support of large-scale data and does not work well in some scenarios, such as those with extreme camera shooting angles. Thus, we propose UEGazeNet, which detects landmarks and obtains the gaze angle from the landmarks. Our method requires relatively few training data and can fit complete eye information from incomplete eye images, which means that the gaze can be detected when the camera shooting angle is extreme or when the eyes are partially obscured. In addition, we also propose another network-UEGazeNet*, which similar to the structure of UEGazeNet but can recognize gaze directions with low-resolution images.

\bgroup
\begin{figure}[!htbp]
\centering
\includegraphics[width=3.33in, keepaspectratio]{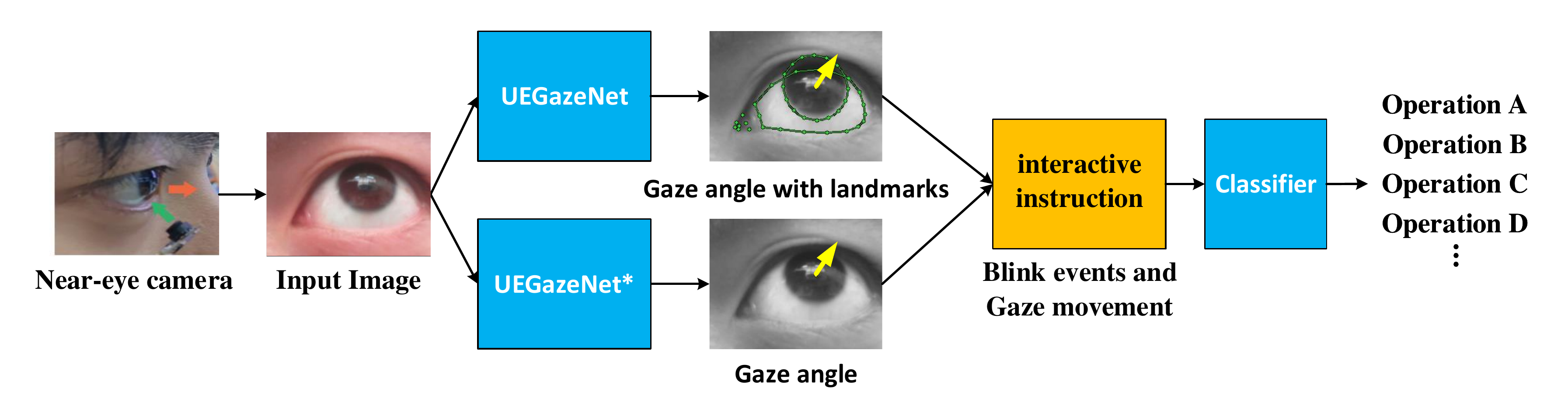}{}
\caption{{Overview of UEGazeNet and UEGazeNet*-two near-eye gaze estimation neural networks and their applications in human-computer interaction.}}
\label{Overview}
\end{figure}
\egroup

Moreover, we design an HCI method based on real-time gaze direction tracking. The traditional HCI methods involve directly controlling cursor movements and switching between interaction events through gaze times or through a blinking action. For example, one existing method allows users to slide their gaze back and forth between the center of the screen and the four corners of the screen to move or rotate a camera [24] [26]. Additionally, dividing the screen into several areas, treating the areas as points, and having users gaze at these points in a specific order to draw gestures is also feasible [25] [27]. However, these approaches require gestures to be mapped to the entire screen, and accuracy depends on gaze tracking. In contrast, our HCI method is based on a gaze gesture classifier that can detect the attention of the user through their gaze gesture. When users use HMDs, they can accomplish tasks well even if they are not in a stationary, stable state, such as when they are walking or performing quick eye movements.

In addition, we establish a gaze-tracking dataset that contains data from 10,200 gaze trajectories collected from 15 individuals using our gaze-tracking method. To ensure diversity, we create random transformations of standard patterns; the participants supply gaze trajectories based on a displayed indicator map, which avoids the problem of participants always using similar gaze tracks.

To summarize, this work has three main contributions. First, we present an effective HMD-based gaze-tracking neural network called UEGazeNet, which can detect landmarks and obtains the gaze angle from the landmarks. Second, we develop a highly robust, flexible and fast-operating HCI method based on the classification of gaze gestures. Third, we design a gaze gesture dataset that is used to train the classifier for HIC. Our HCI approach is faster than regular gaze interaction and very suitable for HMD devices. The rest of the paper is organized as follows. Section 2 provides a review of eye-tracking methods and eye-tracking datasets. Section 3 provides a detailed description of our method, followed by experiments in Section 4 and discussions in Section 5. Finally, Section 6 concludes the paper and proposes further research areas.


 

\section{Related Work}

\subsection{Eye-Tracking Methods}

The process of eye tracking always includes two steps: eye detection and gaze estimation. For eye detection, there are two main methods: those based on shape and those based on appearance. In shape-based methods, the location of the eyes is decided by voting [30] or matching [31] [32] geometrical eye shapes such as the edge shape of the iris or pupil [33], [34], [35], [36]. Generally, these methods require a priori model to judge shape complexities. When the facial posture changes significantly or the image has low resolution, there are few features around the eye area. In this case, the corners of the eyes, eyebrows and other parts of the head can be used to detect eyes. Moreover, the corner of the eyes and the head contour can also be used to constrain the target area [37], [38], [39], [40]. The appearance-based approach uses the appearance of various detection characteristics such as the original color distribution [41], [42], [43], [44], [45] or the distribution after filtering [46], [47], [48], [49], [50].
The gaze can be estimated using either a model-based approach [51], [52], [53], [54] or an appearance-based approach [55], [56], [57], [58]. Model-based approaches simulate the physical structure of the human eye and typically consider physiological behaviors such as eyelid movement. The 3D gaze direction is estimated by assuming a sphere or ellipsoid or by modeling the corneal surface. Generally, these approaches can be divided into two methods: corneal-reflection-based methods [59] [60] and shape-based methods [61] [62]. In corneal-reflection-based approaches, the cornea is irradiated by light, and the first reflected Purkinje image is used for feature detection, which can help to estimate the optical axis in 3D space. This approach requires at least one infrared light source. Shape-based approaches are the same as the shape-based method used in eye detection; that is, gaze estimation is further performed in 3D space based on the eye detection results. However, these approaches rely on measurement information and hardware calibration and require other relevant information, such as camera and monitor position. In addition, they seldom achieve accurate results near the edge regions of the cornea.
 
In contrast, appearance-based approaches calculate the extracted features using regression; thus, these methods do not require camera or geometric calibration. Instead, they directly map an image to the gaze direction. Appearance-based approaches can be divided into parametric forms such as polynomials [63] [64] and nonparametric classifiers such as neural networks [29] [65]. The former, which constitutes the usual practice, obtains the gaze direction through polynomial regression on the dark-bright pupil features generated under the infrared light source. The latter learns a mapping between the two from a large number of "eye image-gaze direction" data. These approaches can implicitly extract the relevant characteristics used to estimate individual changes and identify items of concern, and they do not require scene geometry or camera calibration. Nevertheless, the costs involved in collecting an appropriate dataset cannot be ignored, and these methods generally do not respond well to changes in head pose.

\subsection{Eye-Tracking Datasets}

Several eye-tracking datasets have been developed in recent years, and these datasets can be further classified as real-data-based datasets [29], [66], [67], [68] and synthetic-data-based datasets [28], [69], [70], [71]. Real-data-based datasets use cameras to capture images of eyes and obtain their gaze directions; they are often collected under lab conditions [66], [67], [68] and are not completely applicable to outdoor situations. MPIIGaze [29] is a well-constructed dataset collected from recording 15 users' daily laptop use; however, the gaze direction range of this dataset is narrow. It can be used as a verification set and is satisfactory for unconstrained cross-dataset evaluation, but it is still unsatisfactory as a training dataset.

\bgroup
\begin{figure}[h]
\centering \makeatletter\includegraphics[width=3.45in, keepaspectratio]{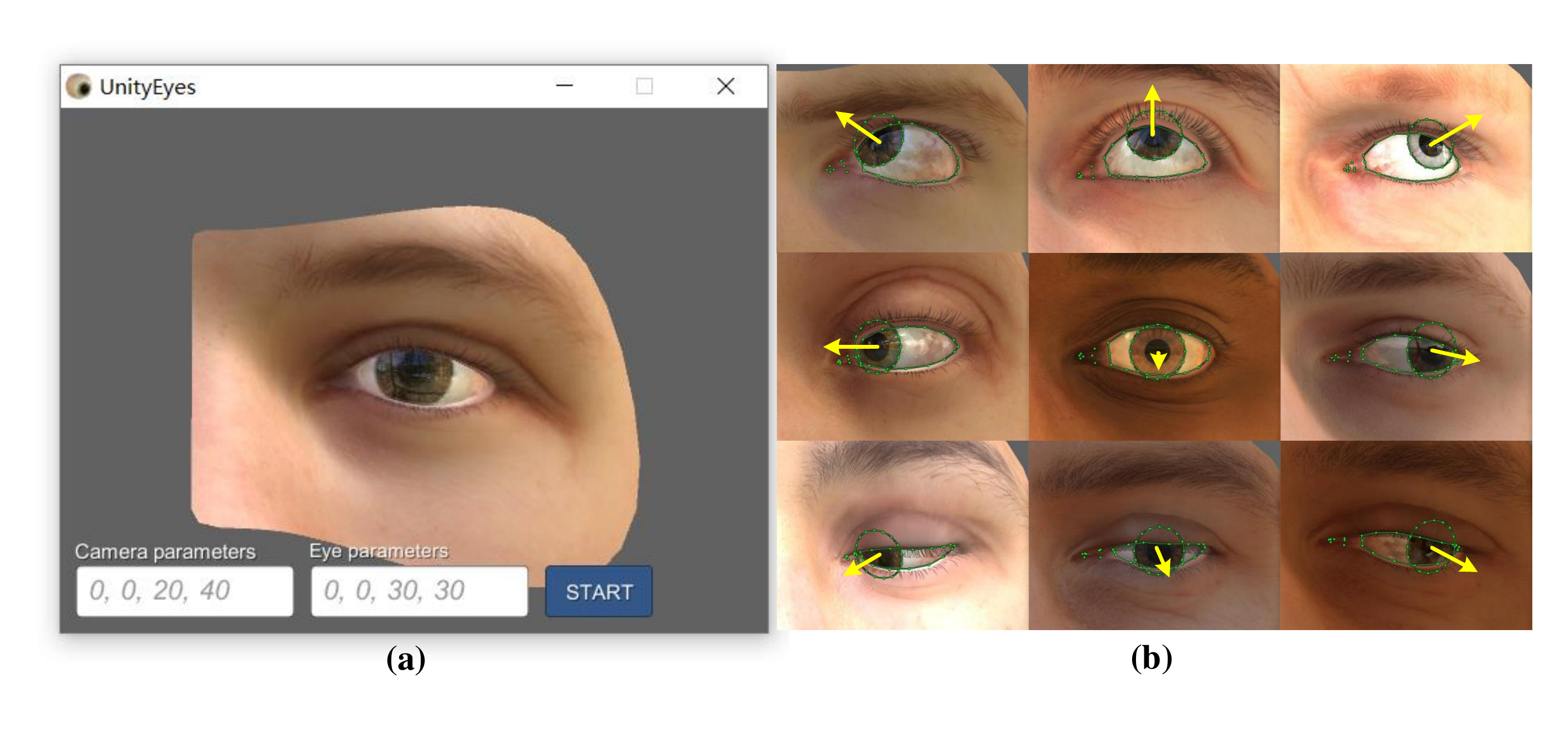}{}
\makeatother 
\caption{{Training data. (a) The software UnityEyes used to generate the training data. We can customize the parameters related to the camera and gaze direction or randomly generate them. (b) The data generated by a simulation model with different lighting and races. Note that the direction of the gaze in the figure is that detected by our method.}}
\label{Training data}
\end{figure}
\egroup

In contrast, synthetic datasets can customize the data following users’ requirements and do not require annotating large numbers of images. UT Multiview [28] uses a combination of virtual and real data and synthesizes data with real data. Due to the gap between the feature distributions in synthetic images and those of real images, learning from synthetic images may not achieve the expected performance. To bridge the gap between a synthetic image distribution and a real image distribution, GazeNet uses a model pretrained on ImageNet to learn from large amounts of data and then trains the resulting model on UT Multiview. Using real data solves the data distribution problem, leading to the proposal of comprehensive learning method [72]. Apple, Inc. proposed using both synthetic data [73] and real unlabeled data to train a model. The approach used synthetic data as the input, and these data can be made to approximate real data through a GAN to enhance the authenticity of the synthetic output while retaining the labeling information through unsupervised learning methods.

\section{Method}
As shown in Fig. \ref{Overview}, our method involves photographing human eyes with a near-eye camera integrated into an HMD to calculate the gaze direction. The perspective affine method is used to map the direction of the gaze to a 3D fixation direction in the target coordinate system. When tracking the gaze direction, blinking is used to switch between interactive events and to start recording the gaze trajectory during the period to identify the user’s operational intent via a classifier.

\subsection{Training Data}

The training data of this study are based on UnityEyes [71] (see Fig. \ref{Training data}), which combines a generated 3D model of the human eye with a real-time rendering frame-work based on high-resolution 3D facial scanning. The model includes eyelid animation that conforms to the human anatomy, and the image reflected in the cornea is a real image. The simulation of corneal curvature and reflection images produces synthetic data for gaze estimation in difficult field situations.

Our method is based on HMD’s near-eye camera, and consequently, we need to customize the range of the head pose distribution to compensate for our method's inability to directly provide head pose information. We collect the data based on the range, and these data are consistent with the image captured by our HMD system. For example, in our method, the camera is in front of and below the human eyes, and it images the human eyes at a certain angle of elevation. We need to increase the number of samples from that angle and similar angles. In addition to head posture, differences in personal appearance significantly impact gaze estimation [29]; therefore, we randomize the appearance when generating data.

\bgroup
\begin{figure}[!htbp]
\centering \makeatletter\includegraphics[width=3.33in, keepaspectratio]{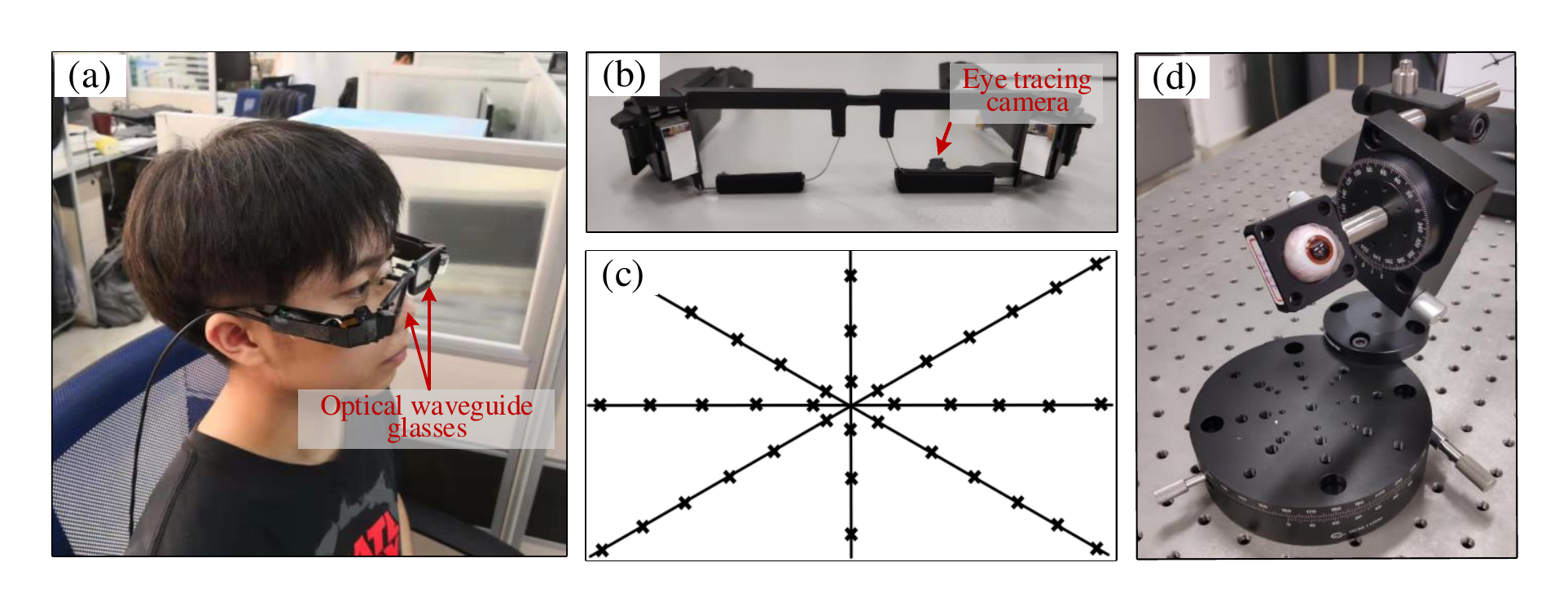}{}
\makeatother 
\caption{{Hardware system. (a) Homemade HMD system for HCI; (b) eye-tracking camera; (c) target board designed for quantitative experiments, which uses (d) the adjustable mechanical device. }}
\label{Hardware system}
\end{figure}
\egroup

\bgroup
\begin{figure*}[h]
\centering \makeatletter\includegraphics[width=7.00in, keepaspectratio]{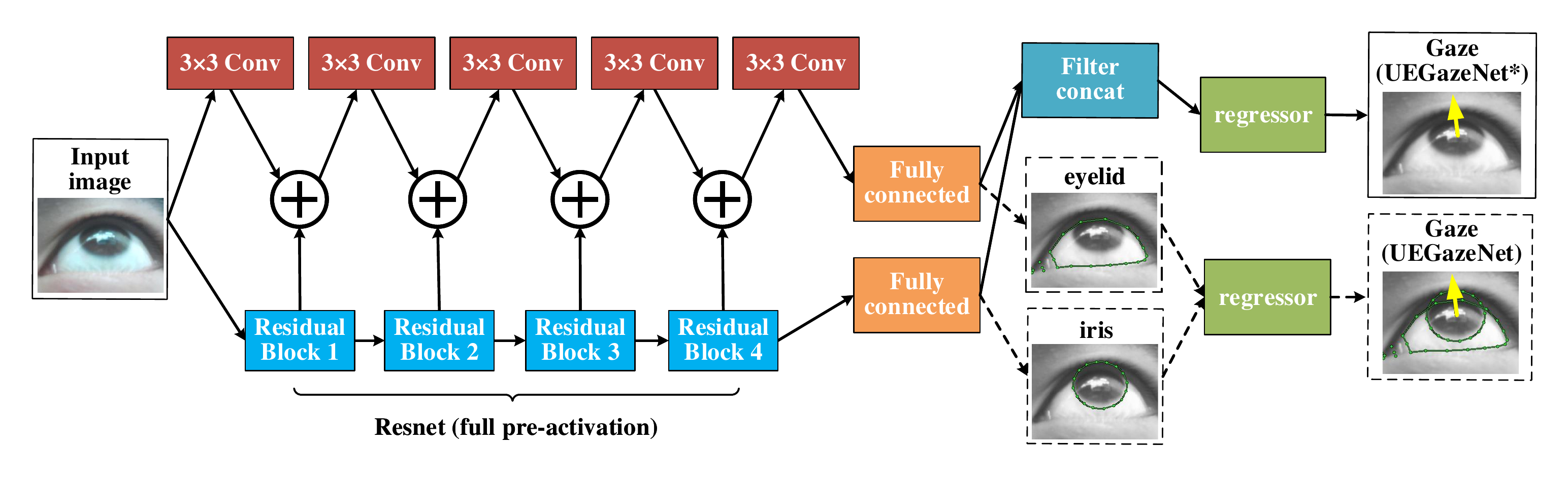}{}
\makeatother 
\caption{{UEGazeNet and UEGazeNet*. Additional layers added on top of ResNet to increase extracted features on different receptive fields while simultaneously detecting iris edges and eyelid.}}
\label{Network}
\end{figure*}
\egroup

\subsection{Hardware Solution}

Our hardware is based on the optical waveguide glasses produced by Lingxi AR Technology CO., Ltd of China (Fig. \ref{Hardware system} (a)). One camera is integrated inside the glasses to capture eye images from a short distance and conduct gaze tracking without requiring extra light (Fig. \ref{Hardware system} (b)). In addition, we use an NVIDIA Jetson TX2, which is a single module based on the AI supercomputing NVIDIA Pascal architecture, for our neural network and 3D rendering synthesis.

\subsection{Gaze Estimation}
\subsubsection{Preprocessing}
Considering that there is no correlation between binocular differences and test results [9], only the data from the left eye are used during model training. To ensure that the eye position is not limited to the middle of the image, we performed the following operations before training.

\begin{enumerate}
            
    \item Randomly enlarge the image: Using the pupil as the center, we randomly magnified the image by a factor of $n$, where $n \in [1, 3]$.
    
    \item Randomly move the image: Using the pupil as the center, we randomly moved $w$ pixels horizontally and $h$ pixels vertically. Note that we ensured the pupil center in the image did not move outside the image boundary as a result, that is, w $\in$ [-x, W-x], h $\in$ [-y, H-y], where W and H are the width and length of the image, respectively, and the pupil center coordinates are (x, y).
    
    \item Randomly rotate the image: Using the pupil as the center, we rotated the image clockwise by a random ($\alpha$) number of degrees, where $\alpha \in [-30^\circ, 30^\circ]$.
    
    \item Randomly reduce the number of image pixels: Our image input size was $256 \times 192$; to support low-resolution input, we randomly performed Gaussian filtering to blur the images.
    
\end{enumerate}

\subsubsection{Neural Network Architecture}
As shown in Fig. \ref{Network}, the neural network structures of our UEGazeNet and UEGazeNet* are partially based on full pre-activation ResNet [74]. However, different from ResNet, the input image synchronously enters a $3 \times 3$ convolutional layer and a residual block, and the outputs of the $3 \times 3$ convolutional layer and the residual block are combined and enter the next $3 \times 3$ convolutional layer. The outputs of both the last convolutional layer and the residual block are connected to a fully connected layer. 

For UEGazeNet, convolution kernels we used for each convolutional layer are respectively 32, 64, 128, 256 and for each residual block including 2 residual units(a residual unit successively including Batch Normalization layers, ReLU layers, Conv  $3 \times 3$ layers, and repeat these three layers once). In particular, the outputs of the two fully connected layers are restricted to extract the landmarks of the eyelid and iris, respectively, which include a total of 55 characteristic points (7 at the corner of the eye, 16 at the eyelid and 32 at the iris). This process is implemented cooperatively by ResNet and the outer nested convolutional layers. Based on the extraction of iris features, the feature points of the eyelids and the corners of the eyes are further extracted. In this way, we can ensure that the feature points of the two parts are relatively independent, while retaining the relationship between shape and position. Thus, our UEGazeNet can achieve good results even under adverse conditions, such as when the eyes are partially obscured, as shown in Fig. \ref{Effects}.

\bgroup
\begin{figure*}[!htbp]
\centering \makeatletter\includegraphics[width=7.00in, keepaspectratio]{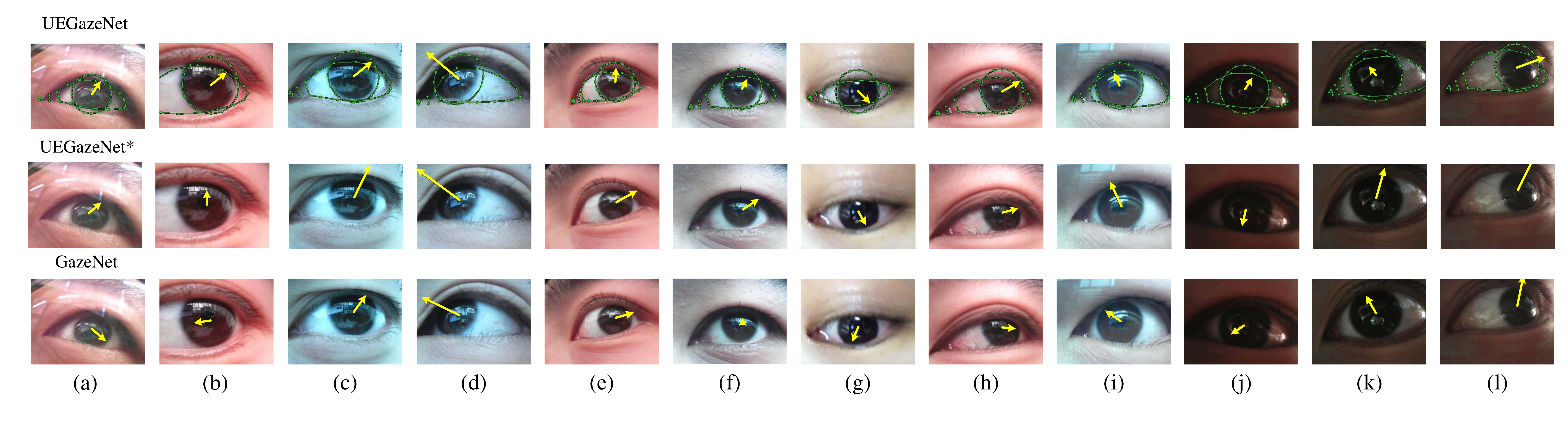}{}
\makeatother 
\caption{{Comparing the effects of UEGazeNet, UEGazeNet*, and GazeNet; all their cameras are on the lower side of the human eye (based on HMD), and GazeNet does not add information about the head posture. (a) Incomplete eyes; (b) (i) (l) a user wearing glasses; (j) (k) (l) a dark environment; (e) the camera is on the nonrental side of the user. }}
\label{Effects}
\end{figure*}
\egroup

For UEGazeNet*, we use respectively 24, 24, 48, 48 convolution kernels for each convolutional layer and 1 residual units  for each residual block. The outputs of two fully connected layers are connected to a filter and then used to directly calculate the gaze direction by regression. In this neural network structure, a series of convolutional layers are connected in sequence outside of ResNet, which allows features to be extracted from the neighborhood of features extracted by ResNet. This approach allows the integration of different receptive features and reduces the impact of the quality of the dataset.

\bgroup
\begin{figure}[!htbp]
\centering \makeatletter\includegraphics[width=3.33in, keepaspectratio]{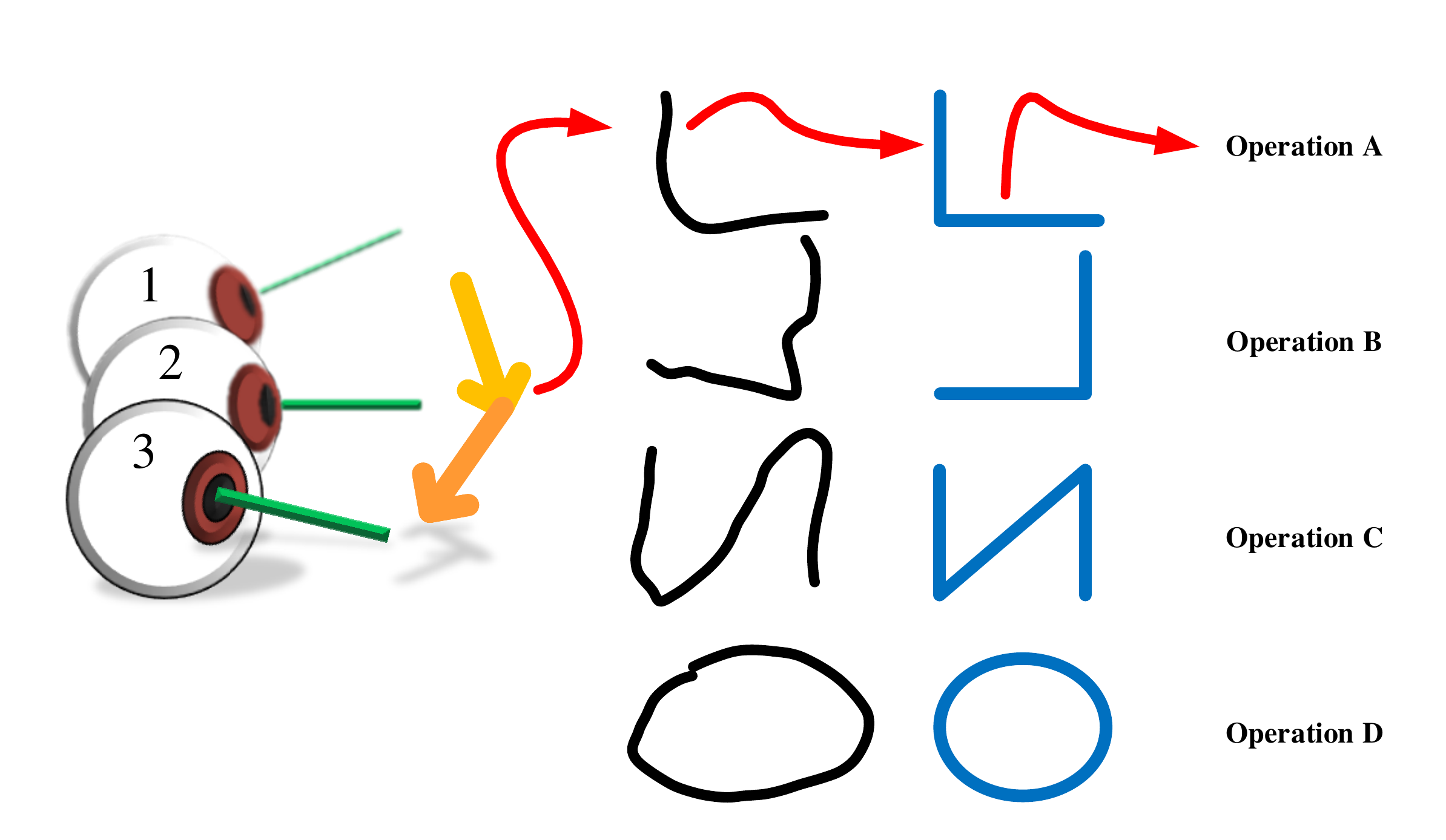}{}
\makeatother 
\caption{{HCI method. In the figure, eyes 1, 2, and 3 show the gaze of an eye at different times, that is, the eye moves from 1 to 2 to 3, drawing a pattern that is recognized by the classifier. The label is also connected as an interface to an operation. }}
\label{HCI method}
\end{figure}
\egroup

\subsection{Human-Computer Interaction}
Traditional HCI methods may be unsuitable when both hands are occupied or in environments in which speech is not an option. At this time, using gaze as an interaction mechanism is an appropriate choice, as gaze behavior is continuous and easy to control. There have long been ways of simulating mouse and keyboard inputs using the eyes, but such interactions are crude and are not sufficiently fast or convenient for HMD applications. Our interaction method uses the user's gaze to "draw" a variety of gestures on the interface, which can be further mapped to a custom operation, as shown in Fig. \ref{HCI method}.

\subsubsection{Collection of the GTgestures Dataset}
We adopted two collection methods: a long-range method and a close-range method. Long-range data collection captures an image of the user’s face through a high-definition camera located in front of the screen to track the gaze direction and record its trajectory for the contrast test. Close-range data collection was based on the HMD; the users start recording data as they use the HMD. We collected a total of 17 patterns for each of the 10 people; data were collected 20 times for each person for each pattern in two batches, i.e., each participant first traced each pattern with their gaze 10 times and then traced the remainder after a break. Finally, we normalized the collected trajectory coordinates using the unit vector of the 3D gaze direction to ensure that the data are usable in different methods.

\subsubsection{GTgesture Classifier}
The collected gestures in the GTgestures dataset consist of two parts. Each of the 10 participants collected 40 patterns (20 for each eye), including a gaze-tracking image with a size of $1,080 \times 1,920$ and a normalized 2D gaze vector. We applied a lightweight CNN network that uses the gaze trajectory image as input to obtain the predicted pattern of category. We use pixel region relationships to resample the images, scale them to $32 \times 32$, and then classify the results via a convolutional layer, a BN layer, two fully connected layers, and the final SoftMax layer.

\section{Experiments}
We conducted a series of assessments of our approach. First, the MPIIGaze’s cross-dataset method [29] is used to evaluate the generalizability of our approach. Then, the eye phantom method is used to evaluate the errors in practical applications, including applications under indoor and outdoor illumination and applications using short-range and long-distance cameras. We also evaluate the GTgestures dataset by using our gaze-tracking methods for HCI applications.

\subsection{Cross-Dataset Evaluation}
In this experiment, UnityEyes and UT Multiview are used as the training dataset and MPIIGaze is used as testing dataset. Compared with the application of HMD, the current state-of-the-art eye-tracking methods always capture the human face at a relatively long distance and then extract the eye position image using an eye recognition algorithm so that the image resolution for eye tracking is low. To ensure the consistency of the algorithm comparison, we uniformly scale the data of UnityEyes and UT Multiview to the scale of MPIIGaze ($60 \times 36$ pixels). The loss function is the Euclidean distance between the real gaze direction and the predicted value and the evaluation criterion is the angle difference between the real 3D gaze direction and the predicted value. We trained each network in 15 epochs with a batch size of 256 on the training set using the Adam solver with an initial learning rate of 0.0001 and multiplied by 0.1 after every 5 epochs. We performed multiple training tests for all models and took the average value instead of adopting the results of multiple random tests after a single training iteration, as sometimes the randomly selected subsets have large distribution differences from the test dataset. We first evaluated the performance on the synthetic and real datasets and then tested the head pose information.

\bgroup
\begin{figure}[!htbp]
\centering \makeatletter\includegraphics[width=3.33in, keepaspectratio]{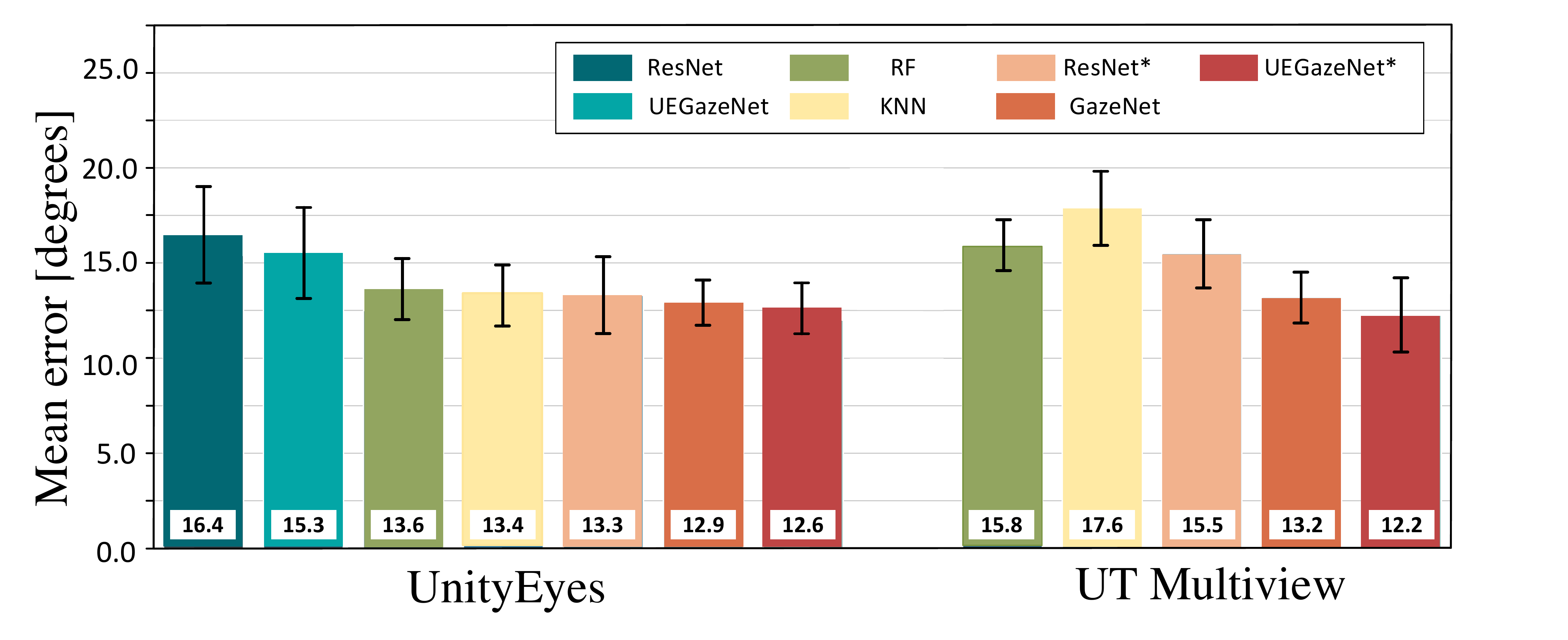}{}
\makeatother 
\caption{{The results of cross-dataset evaluation. We randomly selected 50,000 images and 64,000 images as the training set in UnityEyes (left) and UT Multiview (right), respectively, and then tested them in 45,000 images randomly selected in MPIIGaze. UT does not have fine landmark information; thus, ResNet and UEGazeNet are not compared.}}
\label{The results of cross-dataset evaluation}
\end{figure}
\egroup

\subsubsection{Experimental Results Between the Synthetic and Real Datasets}
Fig. \ref{The results of cross-dataset evaluation} shows the mean angular errors of the different methods, including GazeNet, ResNet (UEGazeNet without the outer nested convolutional layers), ResNet* (full pre-activation), random forest (RF), K-nearest neighbor (KNN), UEGazeNet and UEGazeNet*. Bars correspond to the mean error across all baseline methods in the two datasets, and error bars indicate the standard deviations across each method. As can be seen from the figure, when using the UnityEyes training set, GazeNet can obtain good results even without ImageNet’s pretraining. However, although the error obtained by our UEGazeNet* is minimized, it did not obtain good results when using UEGazeNet. Similar results can also be found for ResNet and ResNet*, which means that, when the image resolution is low, it is more effective to learn gaze direction from the image directly. Moreover, when using the UT Multiview dataset, the best-performing model is still UEGazeNet*, which demonstrates the effectiveness of our network structure. In addition, the performance of RF and KNN showed substantial differences between the two datasets, for which the average errors of the UnityEyes dataset are lower than those of the UT Multiview dataset. The evaluation results of these two methods are largely dependent on the quality of the datasets, which may indicate that the UnityEyes dataset is closer to the real data distribution and can thus achieve a better learning effect.

\bgroup
\begin{figure}[!htbp]
\centering \makeatletter\includegraphics[width=3.33in, keepaspectratio]{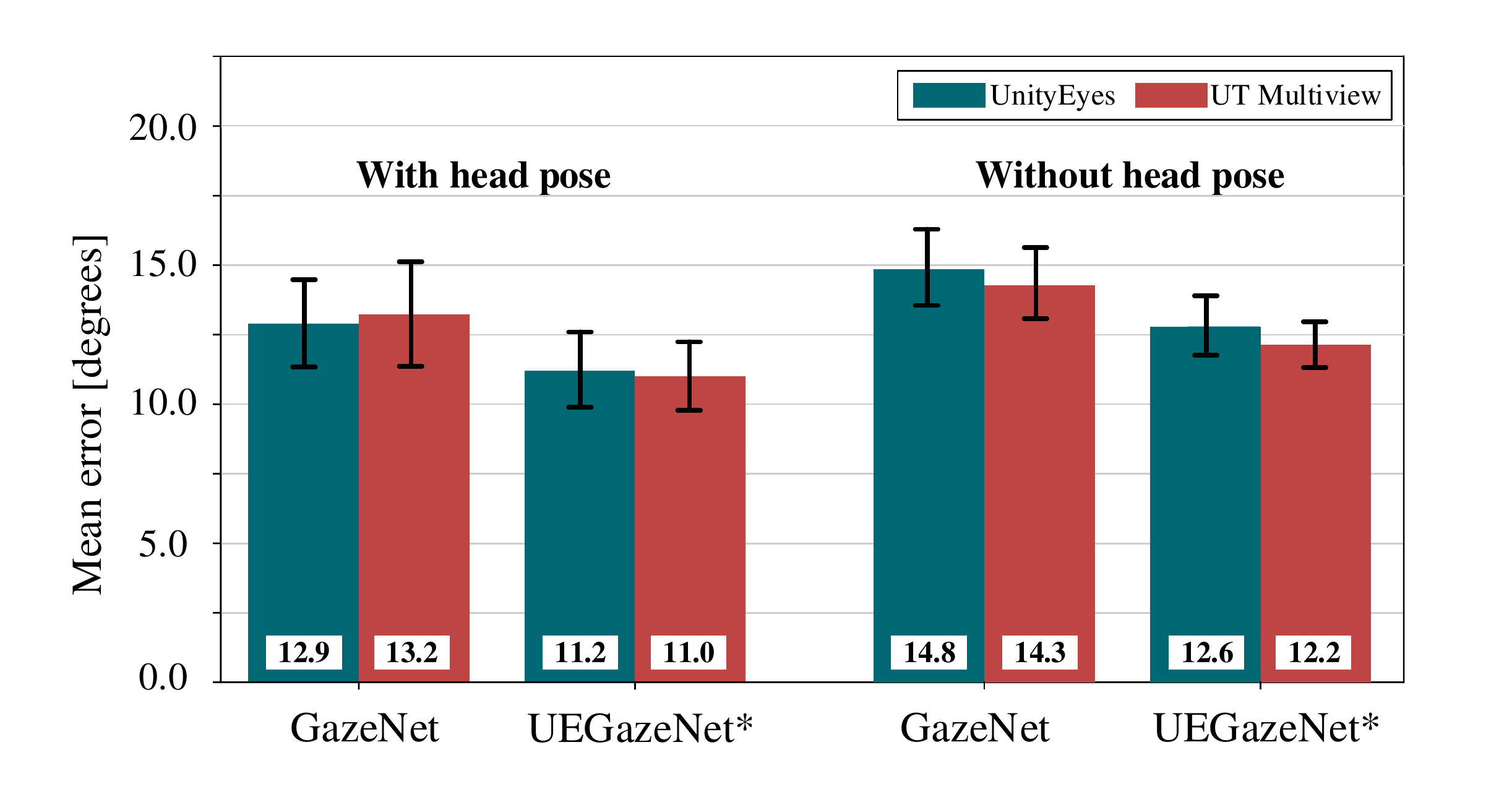}{}
\makeatother 
\caption{{Head pose information. Comparison of the effect of the direct method with that of GazeNet when determining whether to inject header information; both are trained in UnityEyes and UT Multiview. }}
\label{Head pose information}
\end{figure}
\egroup

\subsubsection{Head Pose Information}
To investigate the significance of head posture information, we compared the gaze estimation capability between GazeNet and our UEGazeNet* with and without head posture data. As shown in Fig. \ref{Head pose information}, the average errors of GazeNet and UEGazeNet* are both decreased when adding head posture information, which indicates that the head posture may have a close relationship with gaze direction. Moreover, our UEGazeNet* can achieve a better performance than GazeNet with or without the head posture.

\subsection{Eye Phantom Measurement}
This paper focuses on the near-eye image of HMD’s application; thus, we further evaluate the gaze estimation accuracy by using the eye phantom method [75]. As show in Fig. \ref{Hardware system}.d, we develop a testing stage with a realistic artificial eye and design a mechanical device to adjust its corresponding kinematic model, which enables us to precisely evaluate the eye location of an eye tracker and accurately obtain the gaze direction on the target board (Fig. \ref{Hardware system}.c) by a laser. When we determine the four angles of the target range (top left, top right, bottom left, and bottom right), the gaze directions are transformed to the screen coordinate of HMD by affine transformation. We used UnityEyes as a training dataset to evaluate algorithm performance in different lighting environments and at different image resolutions.

\bgroup
\begin{figure}[!htbp]
\centering \makeatletter\includegraphics[width=3.33in, keepaspectratio]{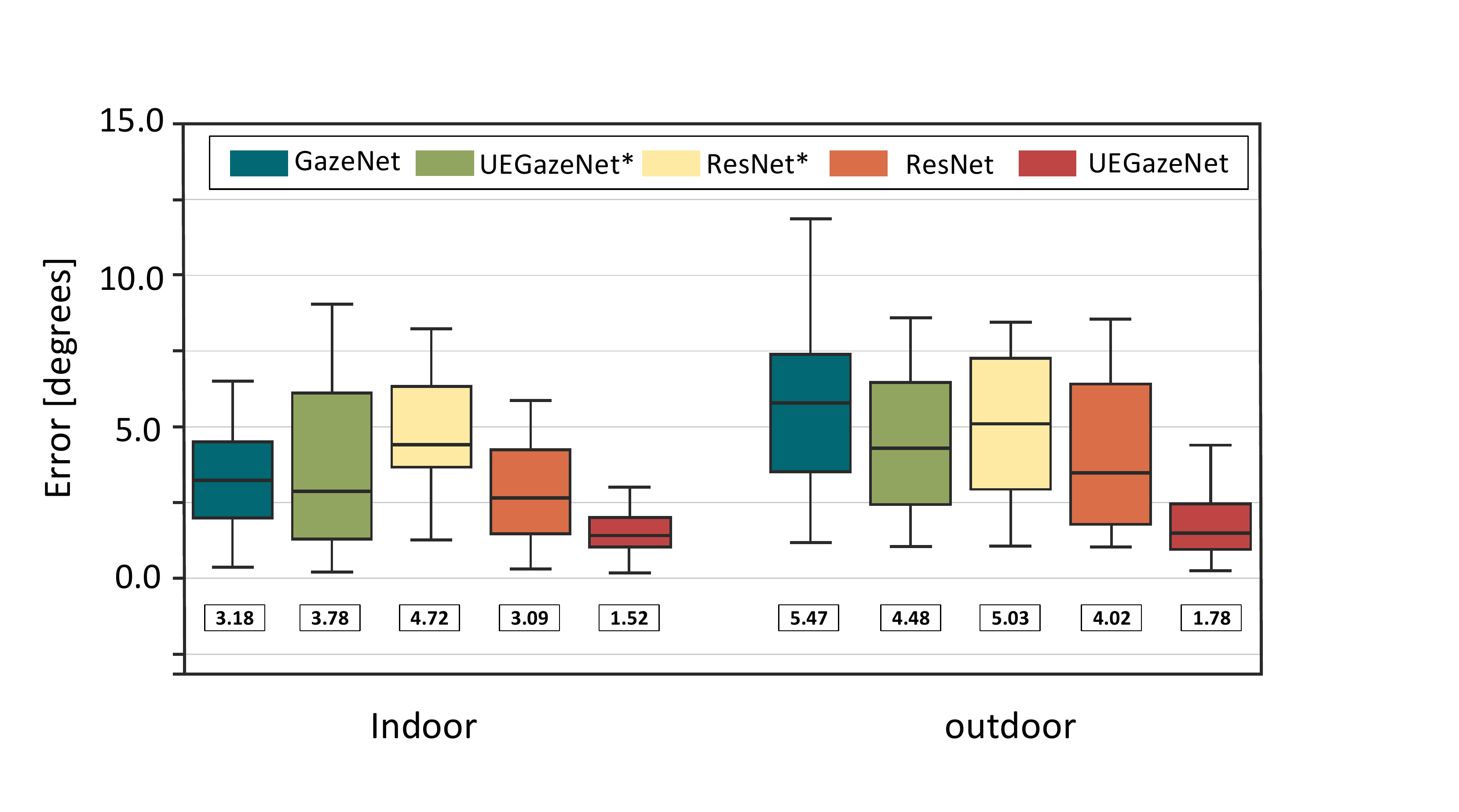}{}
\makeatother 
\caption{{Indoor and outdoor conditions. Indoors, each person is 55 cm from the screen, gazing at the marker points. The result of the estimator is derived, and the angular difference between the two 3D vectors is calculated (left). Outdoor conditions, which are more challenging, are also considered (right). }}
\label{Indoor and outdoor conditions}
\end{figure}
\egroup

\bgroup
\begin{figure*}[!htbp]
\centering \makeatletter\includegraphics[width=7.00in, keepaspectratio]{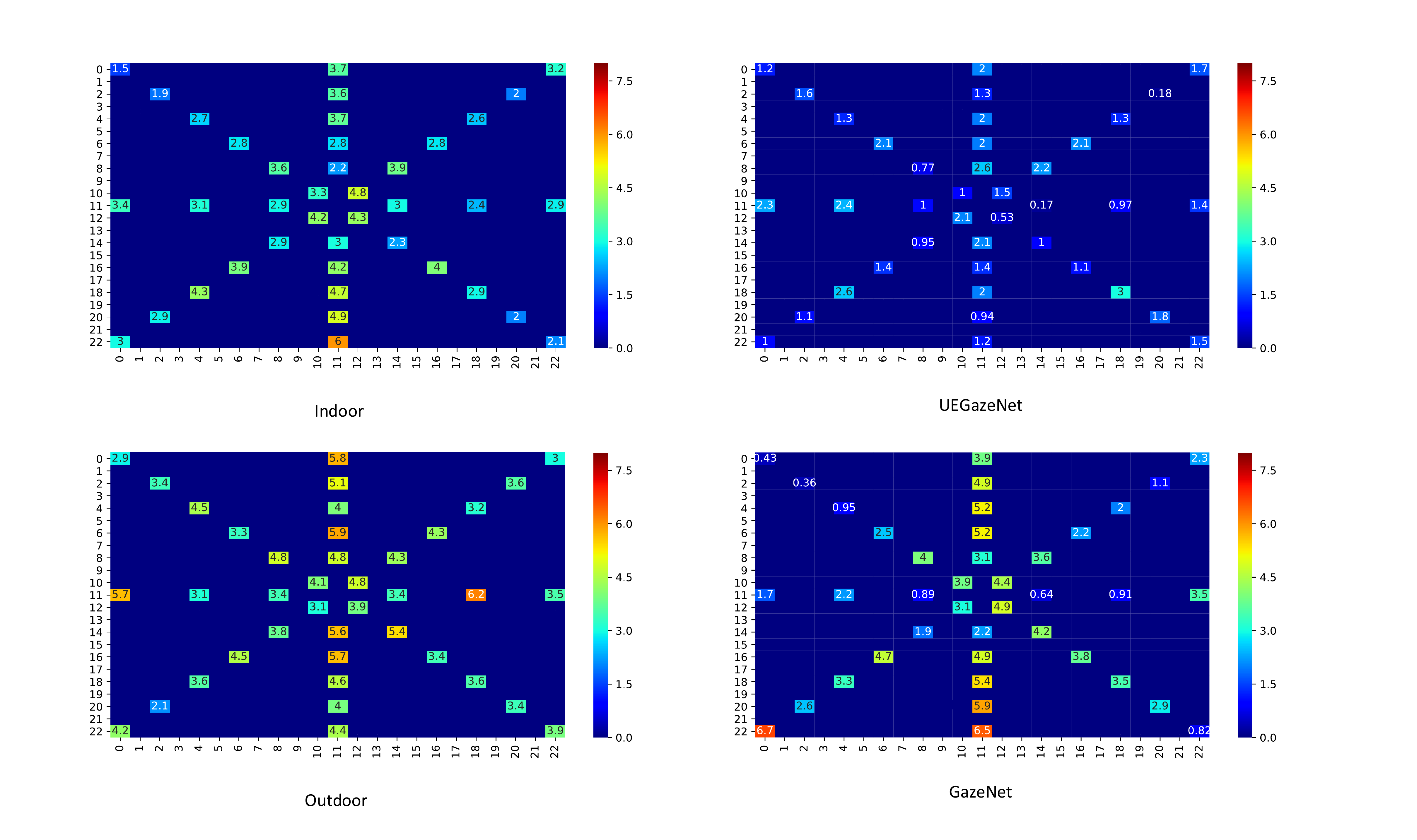}{}
\makeatother 
\caption{{The error distributions measured by the eye phantom.}}
\label{The error distributions measured by the eye phantom}
\end{figure*}
\egroup

\subsubsection{Impact of Light Environment}
In this experiment, the GazeNet, ResNet, UEGazeNet*, ResNet* are used as the baseline methods. To ensure that the training data meets the needs of the model, the size of the training data is set to $60 \times 36$ for GazeNet and $256 \times 196$ for the other methods. Fig. \ref{Indoor and outdoor conditions} shows the error distributions of the different methods when trained on UnityEyes with the near-eye dataset and tested on our designed evaluation system. Bars correspond to the error distribution interval and error bars indicate the standard deviations across each method. As can be seen from the figure, our UEGazeNet shows the lowest error in both indoor and outdoor light environments, with average errors of 1.52 degrees indoors and 1.78 degrees outdoors. In contact, the performance of UEGazeNet* is generally worse on the cross-dataset evaluation, possibly because the eye cannot be fully captured when the gaze direction gradually moves to an extreme angle. Thus, there is a significant effect of landmarks in high-definition images. Moreover, the performance of ResNet (using landmark-based gaze estimation) is not stable, especially outdoors. Compared to GazeNet, the average error of our UEGazeNet is reduced 52.15\% indoors and 67.52\% outdoors, which shows the advantage of our network structure in this challenging test method. 

To analyze the error distribution, we further map the error to the specific location of the target board. Fig. \ref{The error distributions measured by the eye phantom} shows the average error distribution of these methods under indoor or outdoor environments. The indoor error is smaller than the outdoor error and the errors for both are mainly concentrated in the low-middle area because we record the mapping area using four corners. When the gaze angle resolution is low, the middle area is not well distinguished for most methods. However, as can be seen from the figure, the error distribution of our UEGazeNet is relatively balanced, and the errors of GazeNet change significantly when far from the center, which further confirms the validity and stability of our model.

\bgroup
\begin{figure}[!htbp]
\centering \makeatletter\includegraphics[width=3.33in, keepaspectratio]{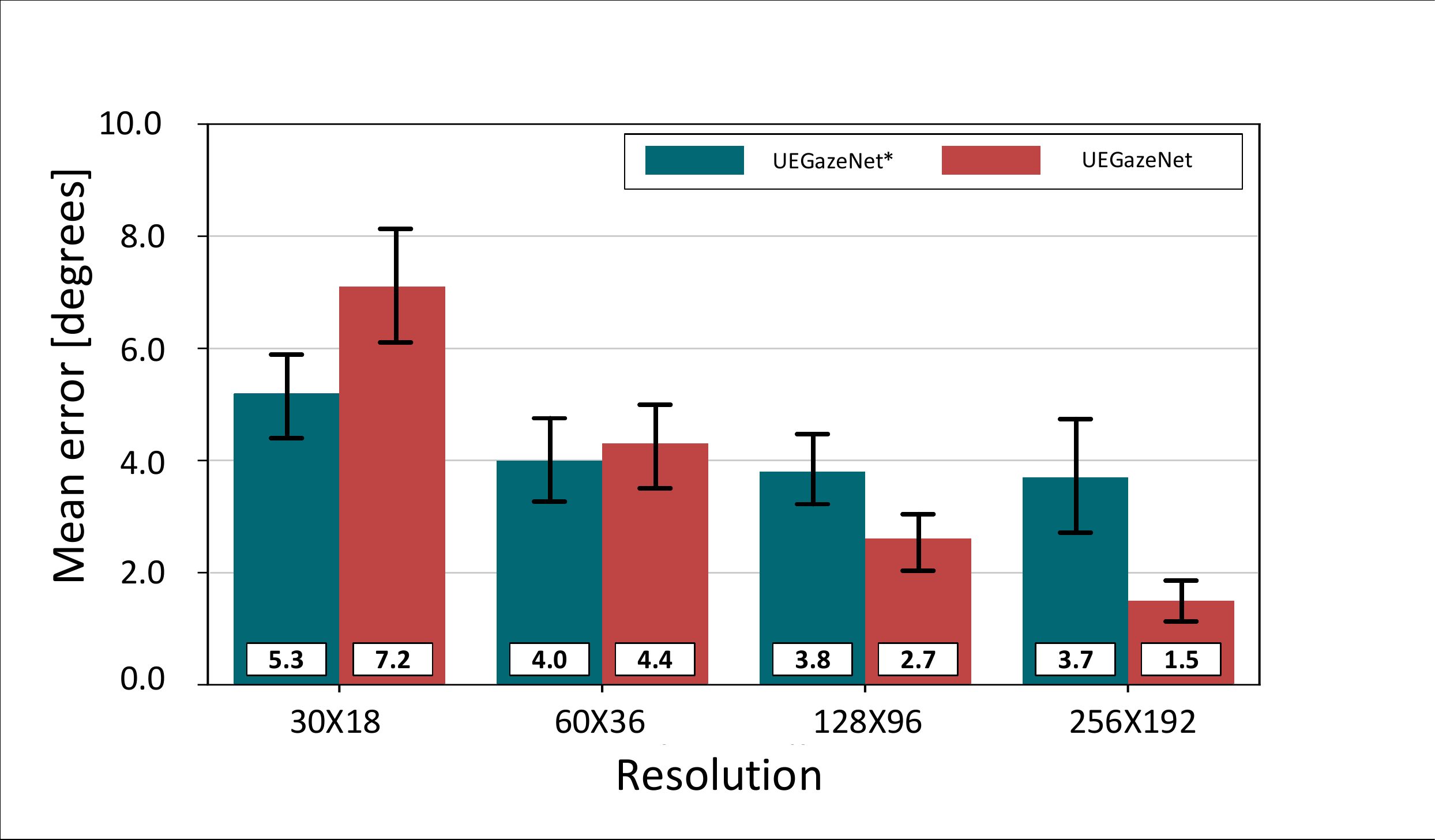}{}
\makeatother 
\caption{{Effects of resolutions. Since our direct method requires high-resolution images (relative to $60 \times 30$), we should explore the effects of resolution on our approach. }}
\label{Effects of resolutions}
\end{figure}
\egroup

\subsubsection{Impact of Resolution}
For different evaluation methods, the results obtained by our two networks are quite different. Thus, we also evaluated our UEGazeNet and UEGazeNet* under different training data sizes of indoor environments. As shown in Fig. \ref{Effects of resolutions}, image resolution has a major impact on gaze estimation performance, especially for UEGazeNet, which relies on landmark detection. When the resolution is low, it is often difficult to find accurate feature locations, which makes it impossible for the gaze estimator to obtain correct features.

\subsection{Evaluation of GTgestures}
The GTgestures dataset contains 17 easy-to-implement patterns that can be divided into 5 categories. We first analyzed the accuracy of the results obtained by different classifiers using our GTgesture dataset. Then, we tested the accuracy of real HCI by using our HMD device. Finally, we evaluated the time spent by different people learning to use interactive patterns. 

\bgroup
\begin{figure}[!htbp]
\centering \makeatletter\includegraphics[width=3.33in, keepaspectratio]{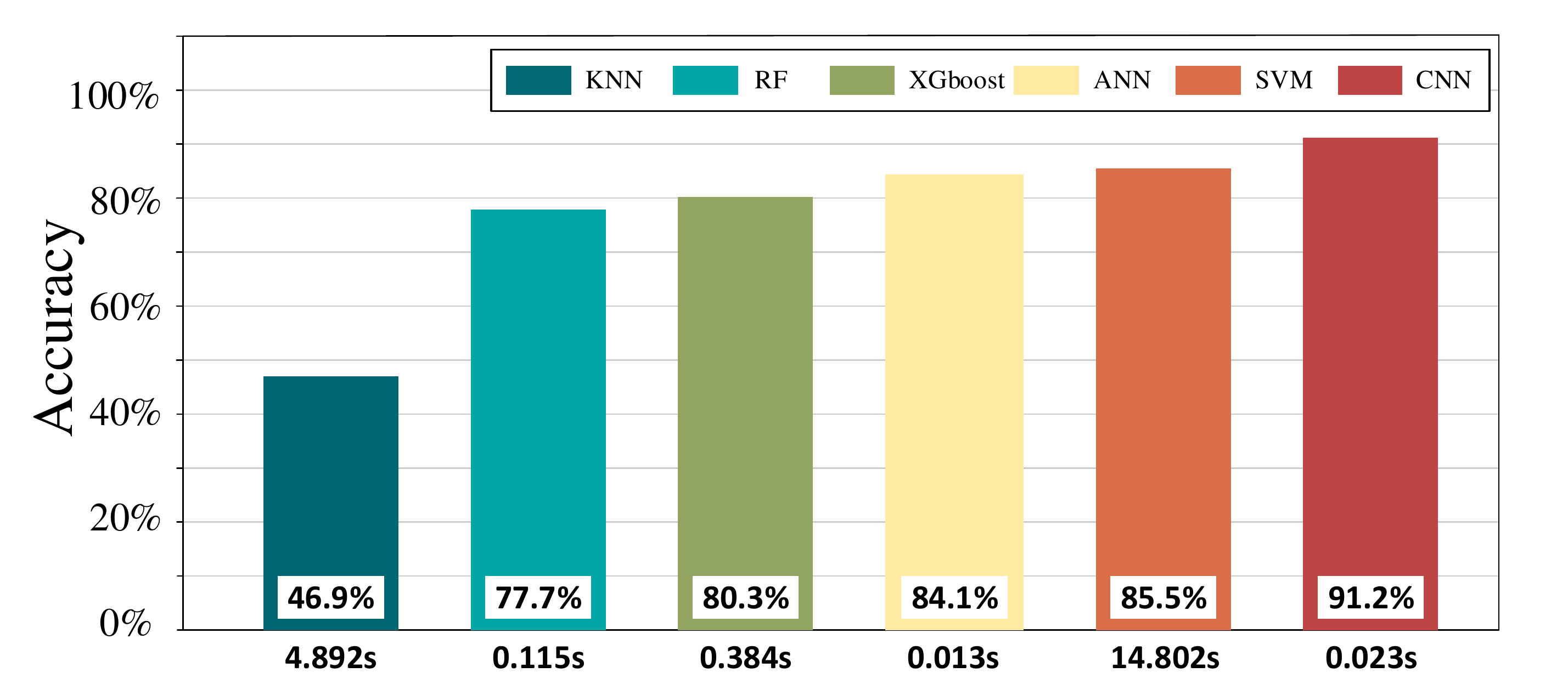}{}
\makeatother 
\caption{{Effects of different classifiers. Different classification methods were used to test the dataset. The image size was $32 \times 32$, and the difference between ANN and CNN is the presence or lack of a convolution layer. }}
\label{Effects of different classifiers}
\end{figure}
\egroup

\subsubsection{Classifier Performance}
We randomly selected 8,160 data of 12 individuals from GTgesture as the training set, and the remaining 2,040 data of 3 individuals were selected as the test set. Some commonly used classifiers were utilized to compare and analyze the results. As shown in Fig. \ref{Effects of different classifiers}, conventional machine learning methods cannot achieve good results, especially for KNN. However, these methods can often get good results for handwritten character recognition, which may indicate that there are some differences between gaze trajectories and handwritten characters. Moreover, CNN can achieve good results for recognition tasks regardless of time and accuracy. In addition, the experimental results also demonstrates  that our GTgesture can be applied to a variety of classifiers and meet the needs of HCI.

\bgroup
\begin{figure}[!htbp]
\centering \makeatletter\includegraphics[width=3.33in, keepaspectratio]{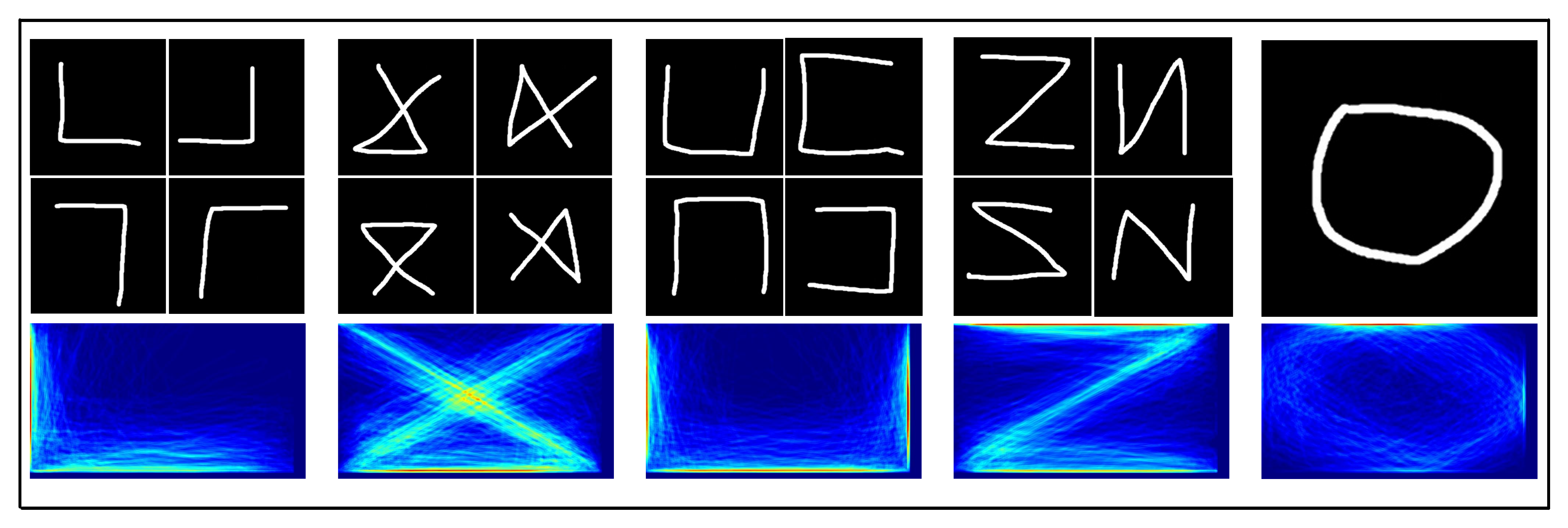}{}
\makeatother 
\caption{{GTgestures. In the figure, there are 17 patterns in 5 categories, and we have evaluated the recognition rate of each pattern. The fifth pattern (circle) has the lowest recognition rate. According to the heat maps, only this pattern produces no clear high-heat area; thus, classification is not effective. At the same time, in general, with a completely closed shape like a circle, people are not easy to draw when drawing patterns.}}
\label{GTgestures}
\end{figure}
\egroup

\subsubsection{HCI performance}
To further evaluate the HCI performance of our method in real applications, we selected 5 individuals who did not participate in GTgestures data collection to conduct experiments using our HMD device. During the evaluation, a pattern randomly appeared on the screen, and the participant traced the pattern with their gaze. The average recognition rate of these 17 gaze trajectories can reach 96.71\%. We further analyzed the recognition rate of each group and found that the first category has the highest recognition rate, which can reach 98\%, while the fifth category has the lowest, reaching only 80\%. Fig. \ref{GTgestures} shows the trajectory probability of the gaze in the form of a heat map. We can clearly see that, among the five patterns, only the fifth (circular) has no high-heat area; that is, the circles drawn by each person are not the same-they differ not only in position but also in the degree of deformation between the traced pattern and the standard template. The other four types have clear high-heat areas following the patterns we designed; therefore, the recognition rates for these patterns are particularly high.

\bgroup
\begin{figure}[!htbp]
\centering 
\includegraphics[width=3.33in, keepaspectratio]{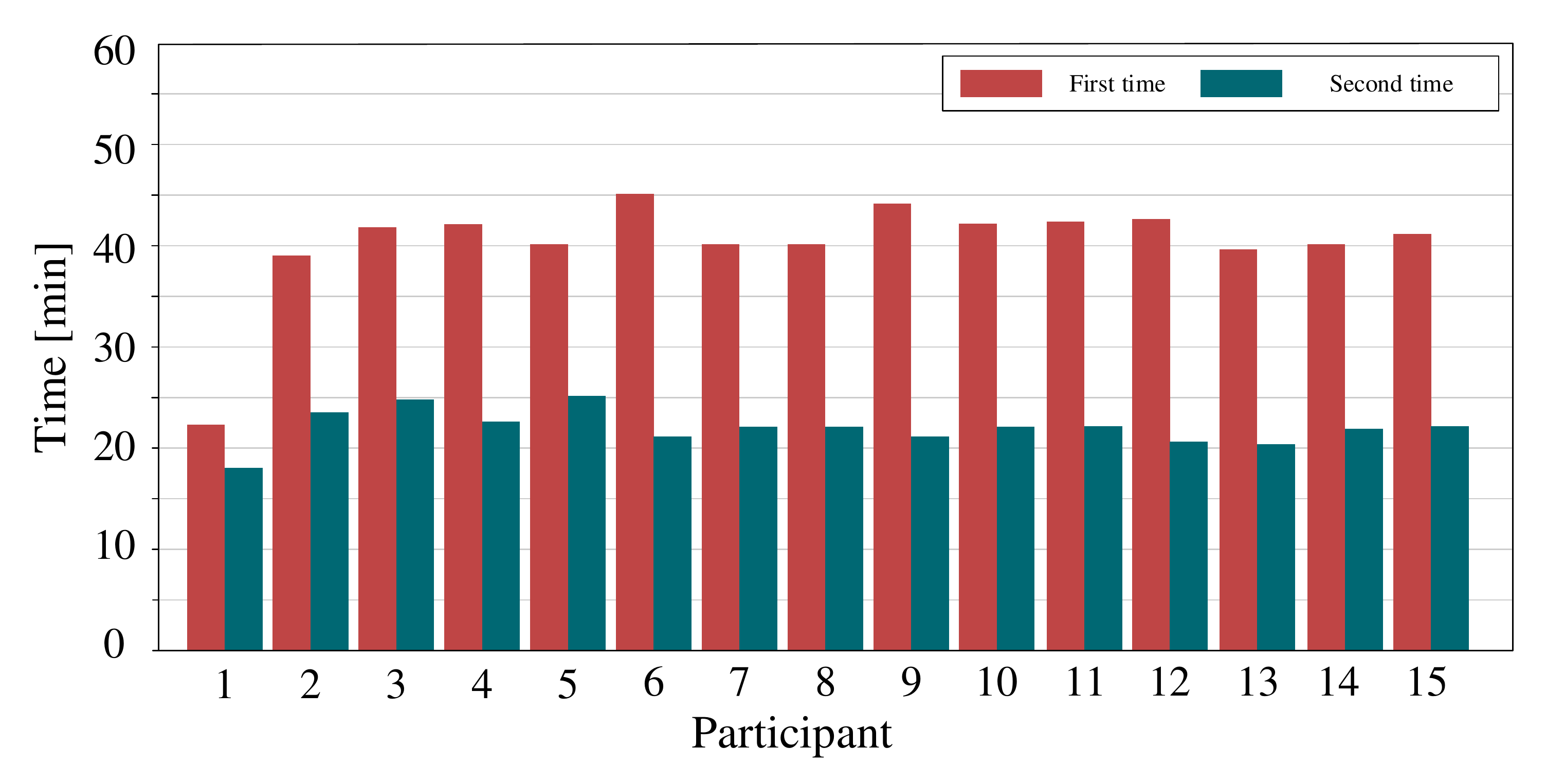}{}
\caption{{Time spent mastering GTgesture. The time spent collecting gaze gesture data from 15 participants. }}
\label{Time spent mastering GTgesture}
\end{figure}
\egroup

\subsubsection{Time Mastering GTgesture}
When collecting GTgesture data, we also recorded the time spent by each participant each time they completed a collection task. We collected data from each person in two batches to analyze the user interactions after their initial experiences and found effective improvements. As shown in Fig. \ref{Time spent mastering GTgesture}, except for participant 1, the time required for the second collection was, on average, half that required for the first collection. Most people require 40-50 min to collect the first batch of data when they are first introduced to this interactive mode, but they require only 20-23 min to collect the second batch of data: decreasing the time required for the task by nearly half. Thus, the gaze gesture is usually used skillfully after the first use.

\section{Discusions}
Unconstrained gaze estimation is the core technology of eye tracking, especially in the HMD field. The relative positions of the camera and the eyes vary from person to person and can shift and rotate with the face when the person wears an HMD. Multiple light sources are often used to deal with these problems [76], and infrared cameras and multiple infrared light sources are used to carry out geometric modeling of the eyeball. However, the use of these devices increases the complexity of system. In contrast, we designed a single RGB camera-based HMD that can achieve unconstrained gaze estimation and HCI through our UEGazeNet or UEGazeNet*. Moreover, instead of using real-world data [29], [66], [67], [68], we generate training data by using the UnityEyes customization according to our requirements [71]. This method can reduce the cost of labeling data and adds data points with extreme angles. As far as the current effect is concerned, we believe that training with synthetic data is an effective approach that is not limited to gaze estimation: it can also be applied in fields such as autonomous driving [77].

Our method uses two neural networks in parallel to extract different features and ensure the correlations between features. UEGazeNet and UEGazeNet* have similar structures but different functions. The experimental results show that UEGazeNet* can consider the global features of the image, which is applicable for low-resolution inputs; meanwhile, UEGazeNet can achieve a good effect through landmark extraction, even under imperfect eye image conditions, which is suitable for close-range detection such as our HMD’s applications.

To improve HCI performance, we do not simply map the gaze to the cursor [24] [26], as the effect of these methods is largely depended on the precision of gaze tracking. Our HCI method is based on a gaze gesture classifier that can detect the attention of the user based on their gaze. In addition, different from previous rule-based classification methods [25], [27], we use a CNN classifier, which does not require very accurate gaze estimation to achieve rapid interaction and can even produce effective interactive operations while in motion.

It is easy to draw inaccurate patterns, especially for users who are just beginning to learn this type of interaction. For example, a new user will take time to think about where to "write" when a trajectory begins or to look elsewhere before the task is complete. We collected a dataset to classify eye patterns without special standardization and accurately detect users' intentions in real time. Based on our extensive evaluation of multiple people and multiple models, the dataset we established is very effective and achieves 96.71\% accuracy. However, when collecting datasets, we found that this interaction mode requires a certain adaptation time, due to people’s attention and habits.

\section{Conclusions}
This paper proposes a gaze-tracking method that uses deep convolutional neural networks to detect the landmarks of eyes and obtains gaze directions from them. This method can learn the relationship between the real human eye and the gaze direction from a small amount of synthetic data. The existence of landmarks allows the model to fit the complete eye information, even when the camera does not capture the complete eye, thus achieving the detection and tracking of the gaze direction. In addition, although this method can adapt to various lighting conditions (including indoor and outdoor environments at different times), its effect also depends on the quality of the input image. Our work demonstrates that it is possible to interact through gaze gestures. Furthermore, we developed the gaze gesture dataset, in which we collected during gaze tracking. For the recognition rate of interaction, we evaluated the 17 patterns in the dataset separately, which proves that the user's intention can be detected even if the gaze-tracking effect is not ideal or the gaze gesture is not accurate. This method is a feasible solution that can be applied to HCI in future HMDs.

\section{Acknowledgments}
This work was supported by the financial support from the National Natural Science Foundation of China (61501101, 61771121), the 111 Project (B16009), and the Fundamental Research Funds for the Central Universities (N171904006, N172410006-2).

\ifCLASSOPTIONcaptionsoff
  \newpage
\fi



%\bibliographystyle{IEEEtran}


\begin{IEEEbiography}[{\includegraphics[width=1in,height=1.25in]{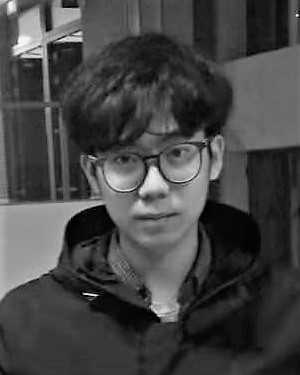}}]{Weixing Chen}
is working toward the BSc degree in Northeastern  university, China. At present, he is an intern at Shenzhen Institutes of Advanced Technology, Chinese Academy of Sciences. His Research experience includes eye tracking , pathological image analysis, low power electrical impedance measurement and non-contact measurements for electrical stimulators. His research interest mainly includes biomedical image processing, pattern recognition.
\end{IEEEbiography}

\vspace{-5ex}

\begin{IEEEbiography}[{\includegraphics[width=1in,height=1.25in]{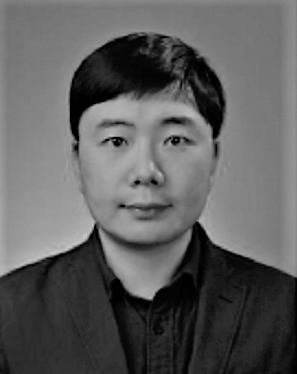}}]{Xiaoyu Cui}
received his Bachelor degrees in Electronics and Information Engineering in 2007 from Shenyang University of Technology and received his Master and Doctor degrees in Biomedical Engineering in 2009 and 2013, respectively, from Northeastern University. He is currently an associate professor in Sino-Dutch Biomedical and Information Engineering School at Northeastern University in China. His research interests include optical imaging and machine learning.
\end{IEEEbiography}

\vspace{-5ex}

\begin{IEEEbiography}[{\includegraphics[width=1in,height=1.25in]{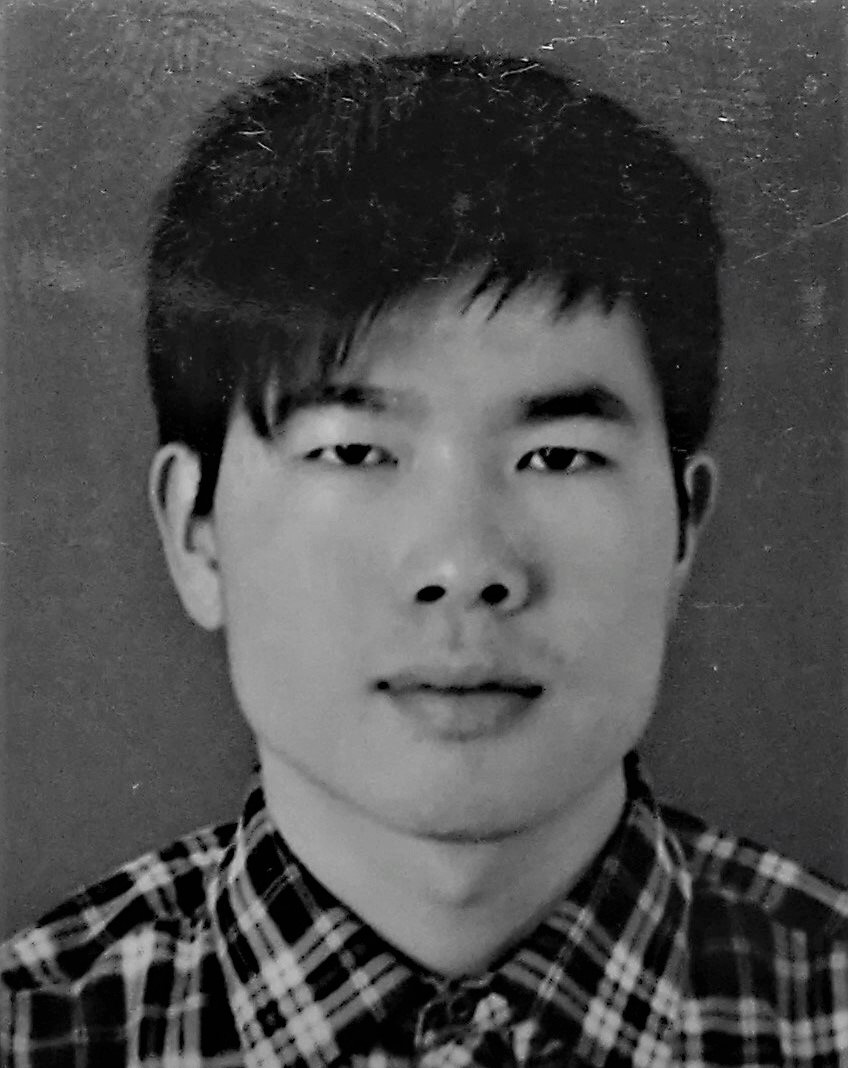}}]{Jing Zheng}
received the BSc degree in machine design from the ShenYang University of Technology,in 2016,and the MSc degree from Northeastern University. His research interests include computer vision,embedded hardware development.  
\end{IEEEbiography}

\vspace{-5ex}

\begin{IEEEbiography}[{\includegraphics[width=1in,height=1.25in]{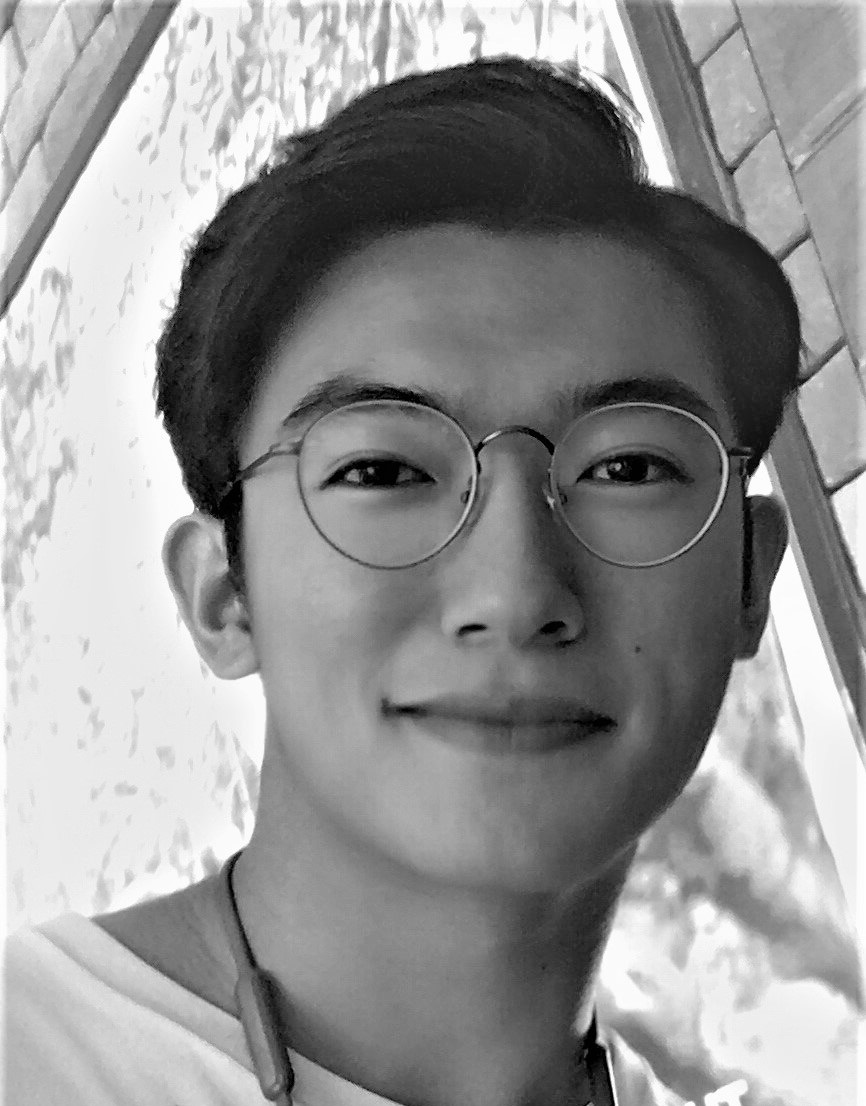}}]{Jinming Zhang}
is working toward a bachelor's degree in Northeastern  university, China. At present, he is an intern at Shenzhen Institutes of Advanced Technology, Chinese Academy of Sciences. His Research experience includes eye tracking and non-contact measurements for electrical stimulators. His research interest mainly includes medical image processing.
\end{IEEEbiography}

\vspace{-5ex}

\begin{IEEEbiography}[{\includegraphics[width=1in,height=1.25in]{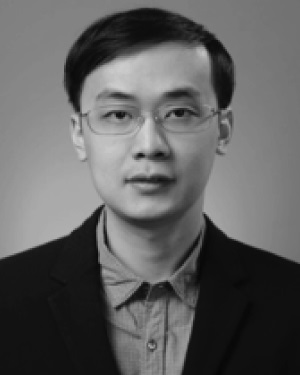}}]{Shuo Chen}
received the B.E. degree in biomedical engineering from Shanghai Jiaotong Univeristy, China, the M.S. degree in biomedical optics from Heidelberg University, Germany, and the Ph.D. degree in biomedical engineering from Nanyang Technological University, Singapore. He is currently an Associate Professor with Northeastern University, China. His research interests include biomedical optical spectroscopy and imaging, noninvasive medical diagnostics, biomedical instrumentation, and biomedical image processing.
\end{IEEEbiography}

\vspace{-5ex}

\begin{IEEEbiography}[{\includegraphics[width=1in,height=1.25in]{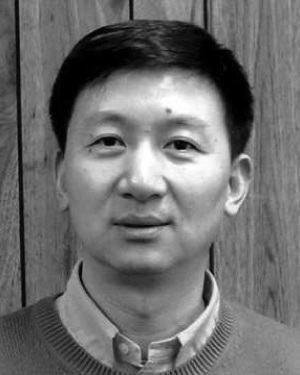}}]{Yudong Yao}
(S'88-M'88-SM'94-F'11) received the B.Eng. and M.Eng. degrees in electrical engineering from the Nanjing University of Posts and Telecommunications, Nanjing, China, in 1982 and 1985, respectively, and the Ph.D. degree in electrical engineering from Southeast University, Nanjing, in 1988. From 1989 and 1990, he was a Research Associate with Carleton University, Ottawa, Canada, focusing on mobile radio communications. From 1990 to 1994, he was with Spar Aerospace Ltd., Montreal, Canada, where he was involved in research on satellite communications. From 1994 to 2000, he was with Qualcomm Inc., San Diego, CA, USA, where he participated in the research and development of wireless code-division multiple-access (CDMA) systems. Since 2000, he has been with the Stevens Institute of Technology, Hoboken, NJ, USA, and is currently a Professor and the Department Director of electrical and computer engineering. He is also a Professor with the Sino-Dutch Biomedical and Information Engineering School, Northeastern University, and the Director of the Stevens’ Wireless Information Systems Engineering Laboratory. He holds one Chinese patent and 12 U.S. patents. His research interests include wireless communications and networks, spread spectrum and CDMA, antenna arrays and beamforming, cognitive and software-defined radio, and digital signal processing for wireless systems. He was an Associate Editor of IEEE Communications Letters and the IEEE Transactions on Vehicular Technology, and an Editor of the IEEE Transactions on Wireless Communications.
\end{IEEEbiography}

%\newpage


%\vfill


\end{document}